# Thermal expansion and Grüneisen parameters of Ba(Fe$_{1-x}$Co$_x$)$_2$As$_2$ – a thermodynamic quest for quantum criticality

Christoph Meingast,[1*] Frédéric Hardy,[1] Rolf Heid,[1] Peter Adelmann,[1] Anna Böhmer,[1,2] Philipp Burger,[1,2] Doris Ernst,[1] Rainer Fromknecht,[1] Peter Schweiss,[1] Thomas Wolf[1]

[1] Institut für Festkörperphysik, Karlsruhe Institute of Technology, 76021 Karlsruhe, Germany.

[2] Fakultät für Physik, Karlsruhe Institute of Technology, 76128 Karlsruhe, Germany.

Thermal expansion data are used to study the uniaxial pressure dependence of the electronic/magnetic entropy of Ba(Fe$_{1-x}$Co$_x$)$_2$As$_2$. Uniaxial pressure is found to be proportional to doping and, thus, also an appropriate tuning parameter in this system. Many of the features predicted to occur for a pressure-tuned quantum critical system, in which superconductivity is an emergent phase hiding the critical point, are observed. The electronic/magnetic Grüneisen parameters associated with the spin-density wave and superconducting transitions further demonstrate an intimate connection between both ordering phenomena.

Unconventional superconductivity often occurs in the vicinity of a magnetic quantum critical point (QCP), i.e. a magnetic instability at which a classical magnetic phase transition is suppressed to zero temperature by a tuning parameter such as doping, pressure or magnetic field. Examples can be found in heavy fermion [1], organic [2], and cuprate [3] materials. In these materials superconductivity usually emerges out of a non-Fermi-liquid (NFL) normal state, which has attracted considerable attention on its own [4]. The NFL state is thought to be a direct consequence of the magnetic fluctuations around the QCP, which have also been linked to the superconducting pairing glue in all these different materials [5] and specifically for the Fe-based materials [6,7]. There now exists mounting evidence also for NFL behavior in the recently discovered iron-based superconductors from resistivity [8,9], NMR [10] and de Haas – van Alphen [11] measurements. Direct thermodynamic evidence of the non-Fermi liquid (NFL) behavior expected near such a quantum critical point in these materials, however, is lacking.

Recently Zhu et al. [12] have shown that the thermal expansion coefficients and the associated Grüneisen parameters offer an ideal thermodynamic probe for locating and classifying a QCP for pressure-tuned systems, since these quantities necessarily diverge at such a point. Further, Garst et al. [13] have shown that, besides the divergence at the QCP, one also expects a sign change of both the thermal expansion coefficients and Grüneisen parameters as one crosses the classical ordering transition away from the critical point.

In this Letter we address the question of possible quantum criticality in the pnictide superconductors through a detailed study of the electronic part of the thermal expansion of Co-doped $BaFe_2As_2$ single-crystals. First, we show that pressure - especially uniaxial pressure - is an appropriate tuning parameter for this system and that there exists a clear accumulation of entropy at optimal doping. Second, we find that the electronic thermal expansivity and derived electronic Grüneisen parameters change sign both at the spin-density wave (SDW) and the SC transitions. This is the expected behavior in a scenario, in which the superconducting dome hides the true critical point. Finally, we observe a remarkable similarity of the electronic Grüneisen parameters associated with the SDW and SC transitions in slightly underdoped and slightly overdoped crystals, respectively, suggesting an intimate connection between both ordering phenomena.

Single crystals of $Ba(Fe_{1-x}Co_x)_2As_2$ were grown from self-flux using pre-reacted FeAs and CoAs powders mixed with Ba. Details of the crystal synthesis are described elsewhere [14-16]. The precise composition of all samples was determined by energy dispersive x-ray spectroscopy and complemented by 4-circle x-ray profile refinement. Dimensions of the crystals were roughly 3 – 9 mm along the a-axis and 0.1 – 0.5 mm along the c-axis. The thermal expansion was measured using a home-built capacitance dilatometer with a typical relative resolution of $\Delta L/L \sim 10^{-8}$ - $10^{-10}$ [17]. All in-plane measurements were made along the [100] direction of the high-temperature tetragonal unit cell; this ensures that the crystals are not detwinned by the small force applied by the dilatometer, and that, therefore, the measured thermal expansion is a simple average of the a- and b- orthorhombic axes. Performing reproducible c-axis measurements proved to be quite difficult due to the large aspect ratios of the crystals, and we therefore present here only the reasonable data. Thermal expansion on detwinned crystals, as well as a detailed analysis of the problems associated with c-axis measurements will be reported elsewhere [18].

Fig. 1 shows the thermal expansion coefficients divided by temperature, α/T, for several Co concentrations x between 0 % and 33 % measured along the a- and c-axis, respectively. Prominent anomalies in α/T are seen at both the SDW/structural and SC transitions, with a clear splitting of the SDW and structural transitions for x > 0 as has been previously reported [8,19,20]. We note that the anomalies along the c-axis are always larger in magnitude and of opposite sign than those along the a-axis, implying larger pressure effects along the c-axis. The a-axis data are however of much higher quality due to the morphology of the crystals, and we will concentrate more on this set of data, since we are interested in the detailed temperature and doping dependences of the thermal expansion. In Fig. 1 we have plotted α/T versus T, because we will focus on the electronic/magnetic contributions. For a Fermi liquid one expects a constant electronic α/T term ($\alpha^{elec.}/T$), which via a Maxwell relation can be shown to simply equal the pressure dependence of $C^{elec.}/T$, i.e. the normal-state Sommerfeld coefficient $\gamma_N$: $\frac{\alpha_i^{elec.}}{T} = -\frac{\partial\left(\frac{S^{elec.}}{T}\right)}{\partial p_i} = -\frac{d\gamma_N}{dp_i} (i=a,c)$. $\alpha^{elec.}/T$ can be obtained by extrapolating the normal-state data down to T = 0 K, and an inspection of Fig.1a reveal negative intercepts for the x = 11.9 % and 15 % data and a near zero intercept for the x =33 % sample. This is made more quantitative in Fig. 2b, in which the normal-state values of α/T at 8 K are plotted versus x in order to probe the doping dependence of $\alpha^{elec.}/T$ and $\frac{d\gamma_N}{dp_i}$ (see right-hand scale of Fig. 2b). We choose 8 K, because phonon contributions to the thermal expansion are small and α/T can be reliably determined [21]. In order to interpret these data, we show in Fig. 2a our experimental [16, 22] values of $\gamma_N$, which first increase with doping (due to the suppression of the SDW) up to optimal doping, then decrease and pass through a minimum around 20 – 30 % Co, and finally increase again up to a quite large value of 43 mJ/mole K$^{-2}$ for pure BaCo$_2$As$_2$. The sharp maximum in $\gamma_N$ at optimal doping is equivalent to a peak in, or an accumulation of, normal-state entropy available for superconducting pairing. $\gamma_N$ values were also calculated using density-functional theory (DFT) using the virtual-crystal approximation (see Fig. 2a) [23] and are found to exhibit a very similar doping dependence as the experimental one; the absolute values are however roughly a factor of two smaller. Similar factors have been reported in both angle-resolved photoemission data [24] and dynamical-mean-field calculations [25] and have been attributed to moderate correlations. In our calculations, the decrease of $\gamma_N$ in the range x = 5 – 20 % results from a shifting of the hole bands below the Fermi energy. Of significance here is that the doping dependences of $\alpha^{elec.}/T$ or $d\gamma_N/dp_i$ shown in Fig. 2b is very reminiscent of the derivative of $\gamma_N$ with respect to doping, i.e. $d\gamma_N/dx$. This clearly demonstrates that for each Co concentration the effect of uniaxial pressure is directly proportional to additional doping, i.e. doping and pressure are directly related over the whole phase diagram. The opposite signs of $d\gamma_N/dp_i$ for the a- and c-axes shown in Fig. 2b moreover demonstrate that the crucial parameter is the c/a ratio of the lattice parameters [14, 26] and not simply the volume [27, 28]. Equating pressure and doping in the superconducting region of the phase diagram, we find that an increase in doping level by 1 % is roughly equivalent to – 1.4 GPa (+ 0.6 GPa) of uniaxial pressure applied along the a- axis (c-axis). A more detailed discussion of the uniaxial pressure effects will be presented in a separate paper [18].

Having established that uniaxial pressure is a good tuning parameter in the Co-doped Ba122 system, we proceed by studying the temperature and doping dependences of electronic/magnetic contribution to the thermal expansion, $\alpha^{elec.}/T$. To do this, we first need to subtract the phonon contribution. Fortuitously, the electronic contribution vanishes near x = 0.33 for both axes (see Fig. 2b), and these data can therefore be used as a background. We note that the phonon background depends only very weakly on the Co-concentration [16]. The resulting $\alpha_{elec.}/T$ versus T curves are shown in Fig. 3. The first thing to note are the negative (positive) contributions above $T_{SDW}$ or $T_c$, which are practically independent of doping and imply constant positive (negative) values of $d\gamma_N/dp_a$ ($d\gamma_N/dp_c$). Again equating doping with uniaxial pressure implies a linearly decreasing electronic entropy with doping (or pressure) for T > $T_{SDW}$ for x ≤ 15 %. An expanded view of this contribution (see Fig. 3c) shows that it is however not constant as a function of temperature; it grows in magnitude with decreasing temperatures, suggesting NFL behavior extending to at least 300 K. This is seen most prominently for the close to optimally doped sample with x = 6.5 %. For the overdoped samples, nearly constant values of $\alpha^{elec.}/T$ are found below a characteristic temperature T*, consistent with a crossover to a FL-like behavior below T*. It is not clear if this apparent NFL behavior is due to magnetic fluctuations or the consequence of a small Fermi temperature in a semi-metal [29]. The second thing to note from Fig. 3 are the sign changes of $\alpha^{elec.}/T$ near $T_{SDW}$ and $T_c$. For example, $\alpha_{elec.}/T$ for the a-axis changes from negative to positive values slightly above the SDW/structural transitions for underdoped crystals (black and blue curves) and at $T_c$ for overdoped superconducting crystals (red curves). For underdoped superconducting samples (blue curves) there exists a second sign change of $\alpha_{elec.}/T$ below $T_c$, which implies that the SDW/structural order parameters are reduced in the superconducting state as has been previously observed [30,31]. No anomalies are found in the non-superconducting 15 % Co overdoped sample (green curve). Again, the c-axis data (Fig.3b) show similar effects, which are however larger in magnitude and have opposite signs. The sign changes and crossover temperatures for the a-axis are summarized in the phase diagram shown in the upper inset of Fig.4. The depicted behavior, especially the multiple sign changes in the underdoped region, is fully consistent with the expected behaviour at a pressure-tuned critical point, in which superconductivity is an emergent phenomena and covers up the critical behavior. An analogous thermal expansion derived phase diagram is observed in the weakly coupled 1D chains system $(C_5H_{12}N)_2CuBr_4$ [32].

Finally, we discuss the electronic/magnetic Grüneisen parameter $\Gamma_{Grüneisen}^{elec.}$, which provides a measure of the volume/pressure dependences of the energy scales relevant for magnetism and superconductivity. Following [12,13] we simply define $\Gamma_{Grüneisen}^{elec.}$ as the ratio of $\alpha^{elec.}/C^{elec.}$, where $C^{elect.}$ is the electronic specific heat taken from Ref. [16,22]. Fig. 4 shows $\Gamma_{Grüneisen}^{elec.}$ for the a-axis as a function of temperature. Overall the behavior is similar to $\alpha^{elec.}/T$ shown in Fig. 3a. The anomalies at the SDW/structural transitions are, however, much less pronounced in $\Gamma_{Grüneisen}^{elec.}$ than in $\alpha^{elec.}/T$, especially for the x = 0 % data, due to appropriately large heat capacity peaks, which normalize the peaks in $\alpha^{elec.}/T$. This highlights the importance of the anomalous high-temperature contribution relative to the actual ordering transitions for the overall behavior of $\Gamma_{Grüneisen}^{elec.}$. There is also little difference between $\Gamma_{Grüneisen}^{elec.}$ at the SDW and structural transitions, implying that both transitions are governed by the same energy scales (i.e. have the same uniaxial pressure dependences). Interestingly, $\Gamma_{Grüneisen}^{elec.}$ in the undoped state (x = 0 %) is practically zero below the SDW transition, suggesting a quenching of the relevant degrees of freedom associated with the anomalous high-temperature NFL state. With increased Co-doping we find increasing positive $\Gamma_{Grüneisen}^{elec.}$ values below $T_{SDW}$. For example, for x = 2 % $\Gamma_{Grüneisen}^{elec.}$ starts to develop a positive signal at low temperatures (below 50 K), which becomes larger and larger for the 4.5 % and 5.5 % samples, almost masking the anomaly at $T_S$ for the 5.5 % sample. The following picture evolves from our data. The excitations associated with the NFL state above $T_{SDW}$, which are fully quenched below $T_{SDW}$ in the undoped state, recover with increased Co doping, and the expected sign changes of $\Gamma_{Grüneisen}^{elec.}$ [13] are clearly observed for the x = 5.5 and 6.5 % samples. The largest values of $\Gamma_{Grüneisen}^{elec.}$ occur for the extremely underdoped (overdoped) samples, for which $\Gamma_{Grüneisen}^{elec.}$ falls (rises) sharply below $T_c$ down to the lowest measured temperatures of 5 K. This shows that superconductivity at these points is extremely sensitive to (uniaxial) pressure, which must be closely related to a strong variation of the residual ungapped density of states observed with doping found in heat capacity data [16]. Finally, we point out the remarkably similar behavior of $\Gamma_{Grüneisen}^{elec.}$ (x = 5.5 %) below the SDW transition and $\Gamma_{Grüneisen}^{elec.}$ (x = 6.5 %) below $T_c$ (see lower inset of Fig. 4). Both data sets exhibit a sign change (near $T_{SDW}$ or $T_c$) and then increase down to lower temperatures. This strong similarity in the magnitude and shape of $\Gamma_{Grüneisen}^{elec.}$ for both SDW and superconductivity clearly demonstrates that both ordering phenomena are intimately related and most likely originate from the same underlying physics.

In summary, we have presented a detailed study of the electronic/magnetic part of the thermal expansion of Co-doped Ba122 single-crystals from which several conclusions can be drawn. First, our results show that pressure - especially uniaxial pressure - is an appropriate tuning parameter in this system and that pressure can be directly linked to doping. Second, the electronic/magnetic thermal expansivity and calculated electronic/magnetic Grüneisen parameters show many of the features predicted to occur for a pressure-tuned quantum critical system in which the true critical point is covered by the SC dome – namely, sign changes of these quantities at both the SDW and SC transitions, NFL behavior above the superconducting dome and a cross-over to FL behavior for higher doping. On the other hand, although the heat capacity data also show a clear accumulation of entropy at optimal doping, this appears to be not primarily the result of an increase in low-energy magnetic fluctuations, since the DFT calculations, which do not include fluctuations, exhibit a similar doping dependence. Most likely, the maximum in $\gamma_N$ results simply from the gradual suppression of the SDW gap with doping at x = 6 % coupled with the disappearance of the hole bands for higher doping levels, which raises the question of the origin of the NFL behavior seen in the thermal expansion. Finally, we observe a remarkable similarity of the electronic/magnetic Grüneisen parameters associated with the SDW and SC transitions in slightly underdoped and slightly overdoped crystals, respectively. Although one expects roughly the same energy scale for both orders at optimal doping due to similar ordering temperatures, the fact that the volume dependences, as implied by the Grüneisen parameters, are also similar suggests an intimate connection between both ordering phenomena.

The authors thank J. Schmalian for helpful discussion. This work has been supported by the Deutsche Forschungsgemeinschaft through SPP1458.

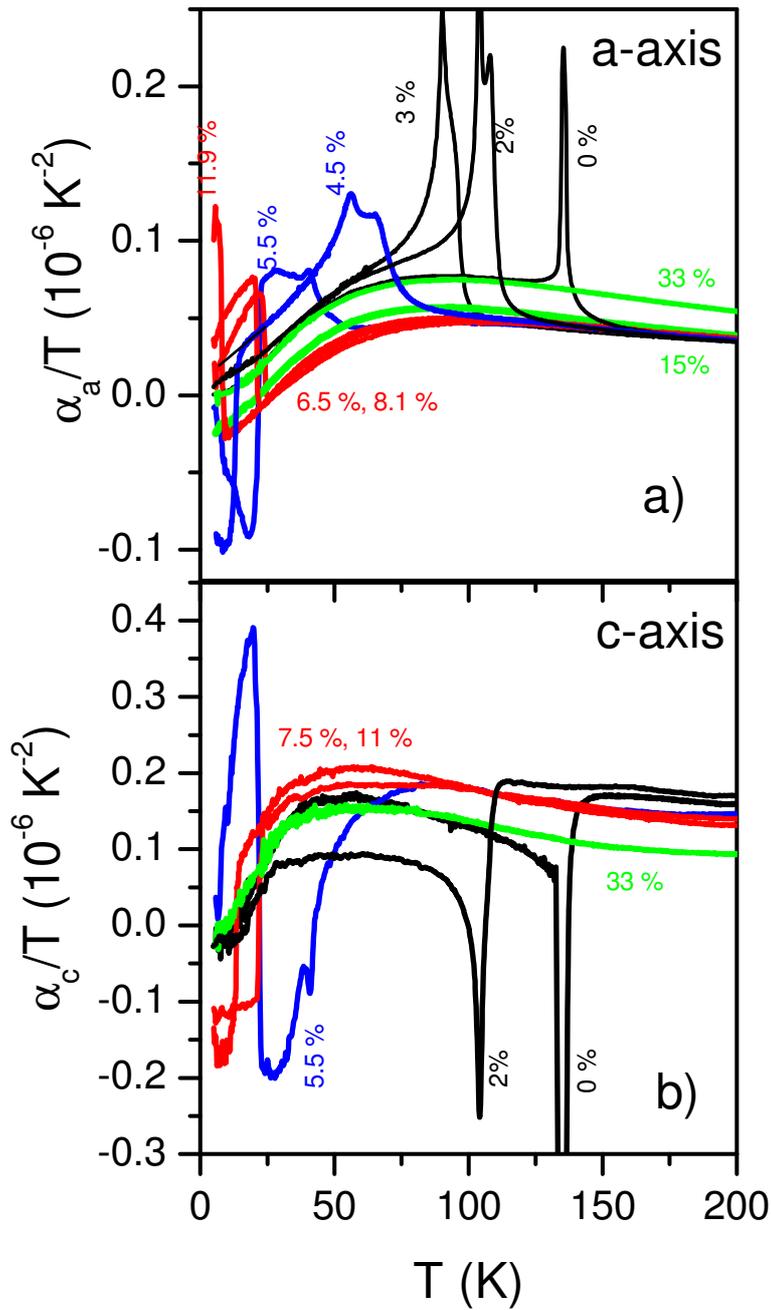

Fig. 1 (color online). Thermal-expansion coefficients divided by temperature, α/T, versus T of Ba(Fe$_{1-x}$Co$_x$)$_2$As$_2$ along a) the tetragonal a-axis and b) c-axis for 0 < x < 33 %. The colors represent the doping levels (black – underdoped, non-superconducting; blue – underdoped, superconducting; red- overdoped, superconducting; green – overdoped, non-superconducting).

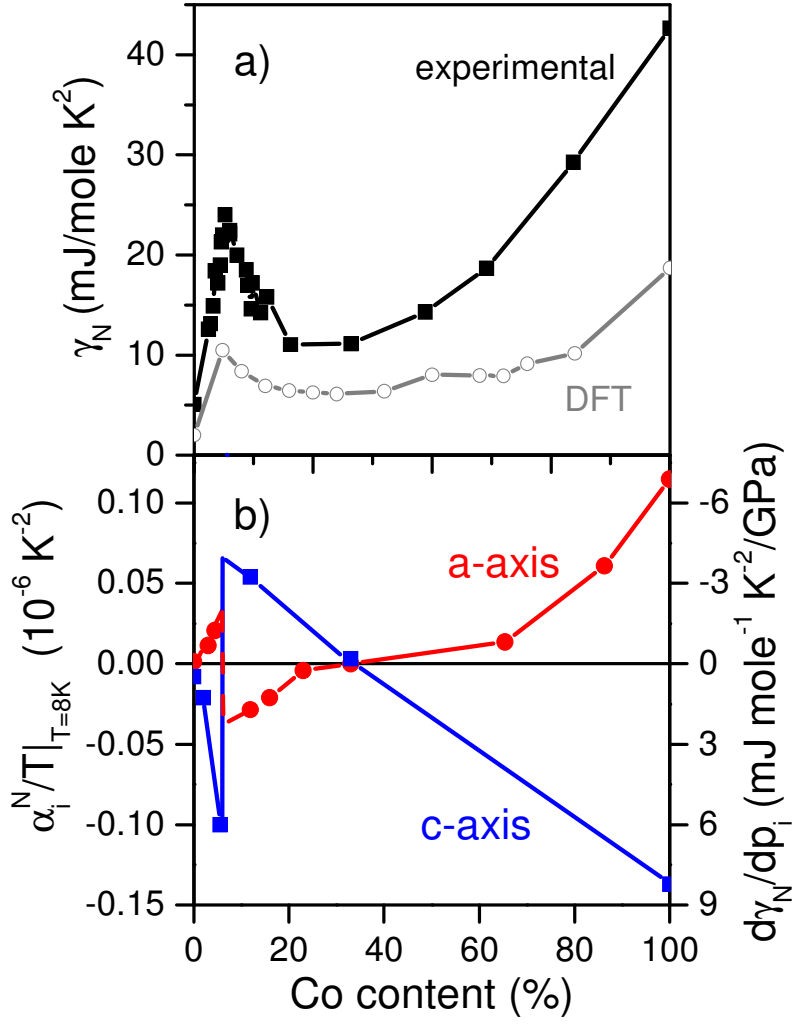

Fig. 2 (color online). a) Normal-state electronic Sommerfeld coefficient, $\gamma_N(x)$, versus Co content derived from experiment [16,22] and theory. b) Normal-state $\alpha/T$ values at 8 K and associated uniaxial pressure derivatives of $\gamma_N(x)$, $\frac{\partial \gamma_N}{\partial p_i} = -\frac{\alpha_i^{elec.}}{T}$ $(i=a,c)$. $\frac{\partial \gamma_N}{\partial p_i}$ and $\frac{\partial \gamma_N}{\partial x}$ derived from a) exhibit similar x-dependences, implying a close relationship between pressure and doping.

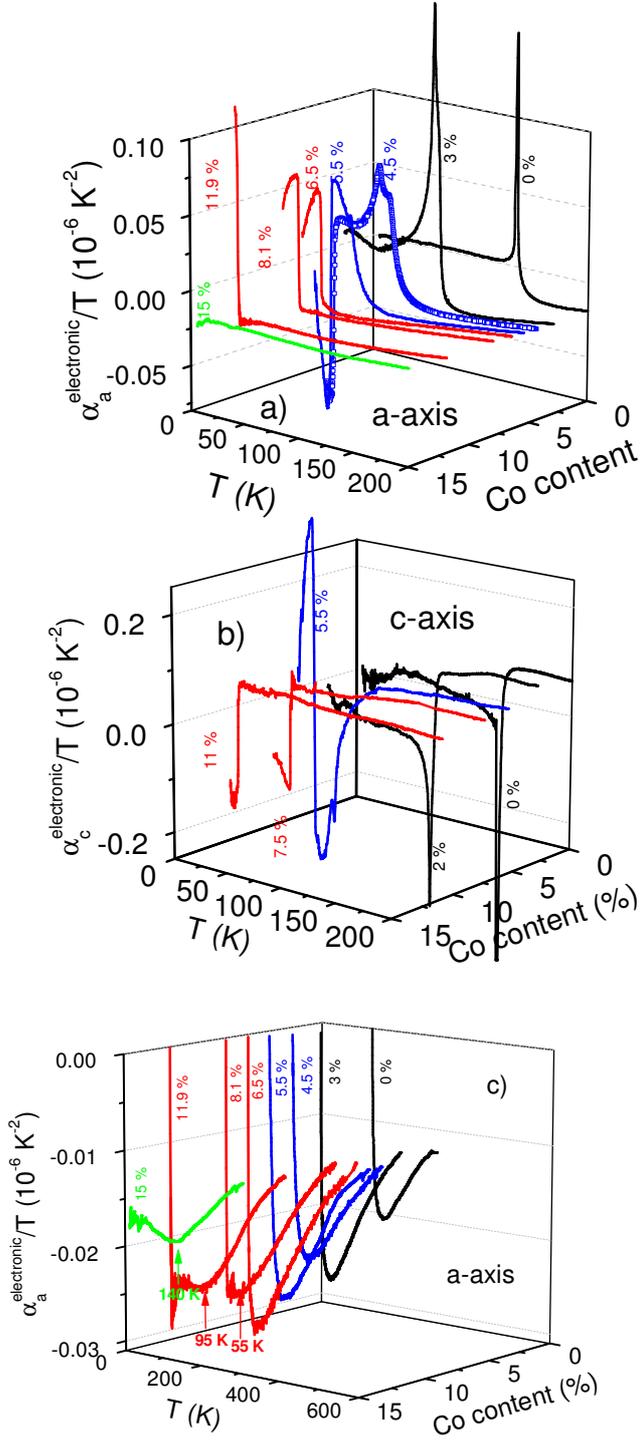

Fig. 3 (color online). Electronic/magnetic contribution of α/T, $\alpha^{elec.}/T$, along a) the tetragonal a-axis and b) the c-axis for 0 < x < 15 % Co. The electronic/magnetic contribution was derived by subtracting the data for x = 33 % Co, which consist mostly of a phonon contribution and a nearly zero electronic term (see Fig. 2b). c) Expanded view of $\alpha_a^{elec.}/T$ showing clear evidence for non-Fermi liquid behavior above $T_{SDW}/T_c$. The arrows mark a crossover temperature, T*, below which Fermi-liquid behavior is observed.

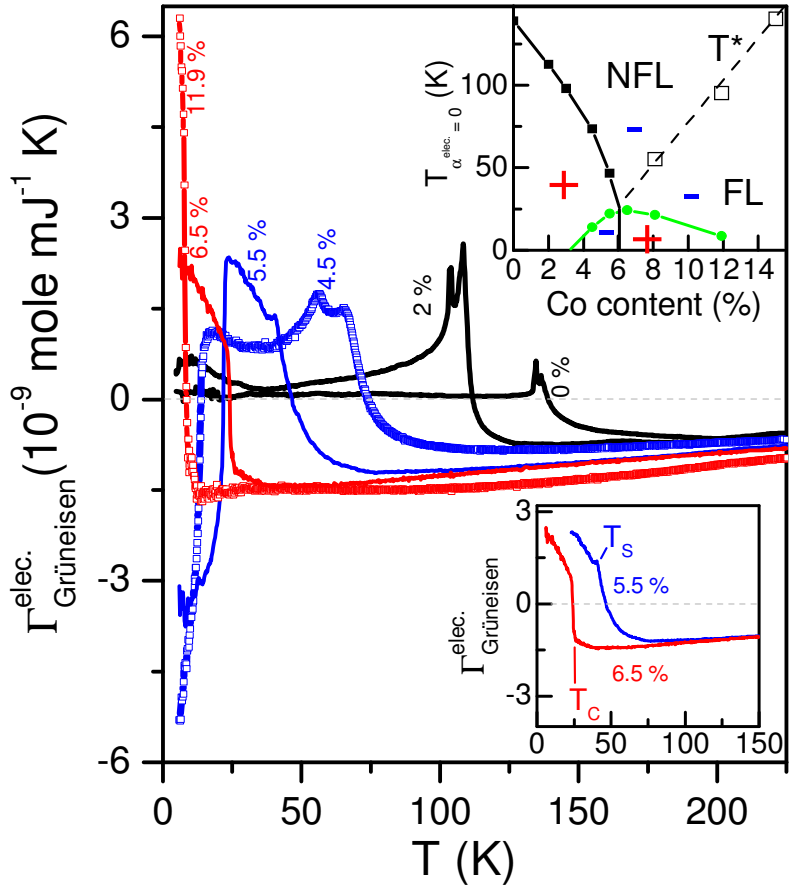

Fig. 4 (color online). Electronic uniaxial Grüneisen parameters, $\Gamma^{elec.}_{Grüneisen}$, for the tetragonal a-axis, derived by subtracting $\alpha^{elec.}_a/T$ data from Fig. 3a by the electronic heat capacity, $C^{elec.}_p/T$, data from Ref. [16,22]. The upper inset shows a phase diagram in which the temperature at which of $\alpha^{elec.}_a/T$ changes sign, as well as the Fermi liquid to non-Fermi liquid crossover temperature T*, are plotted versus x. The lower inset provides a comparison of $\Gamma^{elec.}_{Grüneisen}$ for x = 5.5 and 6 %.